# *Effect of palladium on the microstructure and grain boundary complexions in SiC*

David J. Navarro-Solís, Félix Cancino-Trejo, Eddie López-Honorato*

Centro de Investigación y de Estudios Avanzados del IPN (CINVESTAV), Unidad Saltillo, Av. Industria Metalúrgica 1062, Ramos Arizpe, 25900, México.

*Corresponding author´s e-mail: eddie.lopez@cinvestav.edu.mx

**Abstract**

One of the main challenges in the study of TRISO (Tristructural Isotropic) coated fuel particles is the understanding of the diffusion of fission products through SiC. Among the elements produced inside the uranium kernel, it has been suggested that Pd might enhance the diffusion of other fission products. In this work we have studied the interaction between Pd and SiC. We have observed that as Pd diffuses it can change the chemical composition and microstructure of SiC. Electron Backscattered Diffraction (EBSD) analysis showed that Pd increased the amount of high angle grain boundaries from 47 to 59%. Furthermore, we have observed that as Pd diffused, it changed the composition of SiC by leaving a trail of excess carbon at the grain boundary. This change in localized chemical composition and microstructure suggests a grain boundary complexion transition induced by Pd and a new way in which Pd can lead to faster diffusion routes for other fission products.

## 1. Introduction

The TRISO (Tristructural Isotropic) coated fuel particle consists of a kernel (uranium oxide or oxycarbide) of 300-500 µm in diameter coated with three layers of pyrolytic carbon (PyC) and one of silicon carbide (SiC). The purpose of these coatings is to work as a miniature fission product containment vessel, capable of retaining within the particle all relevant fission products. Among these coatings, SiC is considered the most important as it retains most of the metallic fission products and provides the mechanical strength to the particle.[1-3]

Among all the fission products generated in the kernel, Pd has received particular attention as it is capable of chemically interacting with SiC producing palladium silicides ($Pd_2Si$ at temperatures between 1000 and 1300 °C) and carbon.[4-6] Furthermore, it has been suggested that Pd can enhance the diffusion of other relevant fission products such as $Ag^{110m}$ (a strong $\gamma$-ray emitter), however, it is still unclear how Pd could enhance such diffusion since the original theory suggesting the formation of a solid solution has not been corroborated experimentally, since Pd and Ag have been observed to diffuse separately.[7-10]

Recently, Cancino et al. identified three types of Dillon-Harmer grain boundary complexions[11] in SiC and suggested that the complex diffusion behavior of fission products could be explained through the existence of grain boundary complexions and complexion transitions.[12] Therefore, in this work, we studied the effect of Pd on the microstructure of SiC. We observed for the first time that Pd can induce important changes in composition in SiC at the grain boundaries. This change in composition could be responsible for some previous results found in the literature suggesting that Pd can enhance the diffusion of Ag.

**2. Experimental procedure**

A SiC wafer produced by chemical vapor deposition (provided by Dow Chemicals, no information of the deposition conditions were provided) was used. Small rectangular pieces of SiC ~3x2 mm were placed in contact with Pd powder (Sigma Aldrich, ≥ 99.9% purity, particle size < 1 µm) at 1300°C for 3 hours under argon atmosphere. After heat treatment, the samples were embedded in copper loaded resin and polished up to its cross-section using SiC grinding paper, diamond paste and colloidal silica.

The microstructure was characterized using a scanning electron microscope (Philips XL30-ESEM) and energy dispersive X-ray spectroscopy (EDS). Electron backscattered diffraction (EBSD) was performed using orientation imaging microscopy (OIM) and the TSL-OIM 3.5 program with 20 kV of accelerating voltage, 15 mm working distance, a 70° tilt and a step size of 0.8 mm; measuring more than 50 distinctive grains for better accuracy. Grain size was measured according to the ASTM-E112-12 standard, whereas a minimum adjacent misorientation of <5° was defined for the identification of grain boundaries, affecting approximately 10% of the data.[13] Samples for transmission electron microscopy (TEM) were obtained using focused ion beam milling (FEI-Scios dual-beam). TEM was performed using an FEI-

Talos F200X with four in-column Super-X EDS detectors with a beam current of 300 pA and a collection time of 10 min.

## 3. Results and Discussion

Figure 1 shows that after heat treatment at 1300 °C, Pd was found well within SiC. An EBSD map was taken in an area where Pd already diffused in order to identify possible changes introduced by Pd. Figure 1b shows the inverse pole figure (IPF) and the standard stereographic triangle (SST), in which a mixture of equiaxed and columnar grains with no preferred orientation can be seen. This random orientation before and after Pd interaction was also corroborated by pole figures in Fig. S1. The grain boundary character distribution before and after heat treatment is shown in Figure 2. It can be seen that as-produced SiC grain boundaries were dominated by high angle (>15°, 47%) and Coincidence Site Lattice (CSL) boundaries (36%), with low angle grain boundaries (<15°) accounting for only 17%. However, after heat treatment with Pd, this distribution changed to 59% high angle grain boundaries, 24% CSL and 17% low angle grain boundaries; 12% more high angle grain boundaries than the as-produced material. This change in microstructure was reflected on the average grain size detected as it changed from 10 to 17 µm after heat treatment.

Furthermore, when element maps where taken on the areas where Pd was present, it was possible to identify Pd, Si, and C (Fig. S2). High concentrations of carbon were observed alongside Pd, showing a deficiency of Si in the same areas (Figs. S2 b-d).

This variation in C and Si concentration due to the presence of Pd was corroborated by TEM. Fig. 3 shows the presence of a Pd cluster of approximately 60 nm in length as it diffused through SiC. Element maps show that as the Pd particle traveled through SiC, it left behind an area with silicon deficiency and excess carbon. This change in composition could be due to the chemical reaction between SiC and Pd producing palladium silicides and carbon.[5, 14] This change in composition could be described as a grain boundary complexion transition towards the formation of a nanolayer. The formation of such a layer could also be seen in Fig. S3, where the formation of a nanolayer with a higher concentration of C compared to bulk SiC was also identified (Table S1) between two triple points containing Pd, thus supporting the idea that Pd can induce changes in the grain boundary structure.

The grain growth and the variations in grain boundary character distribution could be produced by the grain boundary complexion transition induced by Pd as it diffused through SiC. It has been previously reported that nanolayer complexions can have a diffusivity of even two orders of magnitude higher than clean, mono and bilayer complexions and that grain boundary complexion transitions towards more disordered structures can induce grain growth.[15] This grain boundary complexion transition could explain the grain growth induced by Pd and irradiation previously reported by O'Connell et al.[7, 16] Therefore, our results suggest that as Pd diffuses through the grain boundary, it can interact with SiC leading to a higher degree of disorder at the grain boundary. This has important implications since Pd might be able to indirectly affect the diffusion of other elements, not necessarily by forming solid solutions as previously suggested, but by rapidly diffusing through SiC while leaving paths for faster diffusion. This result is supported by previous modeling work that shows that even a single Pd atom could induce distortions at the grain boundary.[17]

Overall, it should be stressed that although it is expected that secondary elements such as Pd can play a role in fission product transport, current results suggest that the presence of secondary elements alone (like Pd) cannot fully explain the enhanced diffusion of elements such as Ag in TRISO fuel particles. It has been observed that Pd can be found in coated particles with either high or low retention of fission products,[16, 18] and that despite having a mixture of Pd/Ag in contact with SiC, only the irradiated sample exhibited higher diffusion rates of these elements. This suggests that it is the combination of irradiation, temperature, secondary elements and chemical potential that will control the energy and transformation of the characteristics of the grain boundaries[19, 20] (grain boundary complexions and complexions transitions) and dictate the diffusivities observed at high temperatures

## 4. Conclusions

In this work we observed that Pd can affect the chemical composition of SiC as it diffuses through SiC by leaving areas with a higher concentration of carbon, thus inducing grain growth and the increase of the concentration of high angle grain boundaries. This suggests that Pd can induce a grain boundary complexion transition by forming an amorphous nanolayer. This work suggests a new way in which Pd can enhance the diffusion of other fission products in TRISO fuel particles.

## 5. Acknowledgments

Authors would like to acknowledge CONACYT for the Ph.D. grants awarded to F. Cancino-Trejo and D. Navarro-Solis. Finally, the authors would like to thank Dow Chemicals for providing the SiC wafers.

**List of Figure Captions**

Figure 1. (a) Presence of Pd within SiC and (b) inverse pole figure (IPF) and the standard stereographic triangle (SST).

Figure 2. Grain boundary character distribution (GBCD) of SiC before and after heat treatment with Pd.

Figure 3. (a) HRTEM image of a particle of palladium in a SiC grain boundary; (b) combined Pd and C EDS map; (c) carbon EDS map y (d) silicon EDS map. In (d) is possible to observe a Si deficiency (indicated by the white bars), whereas in (c) is possible to identify the presence of excess carbon.

**List of Figures**

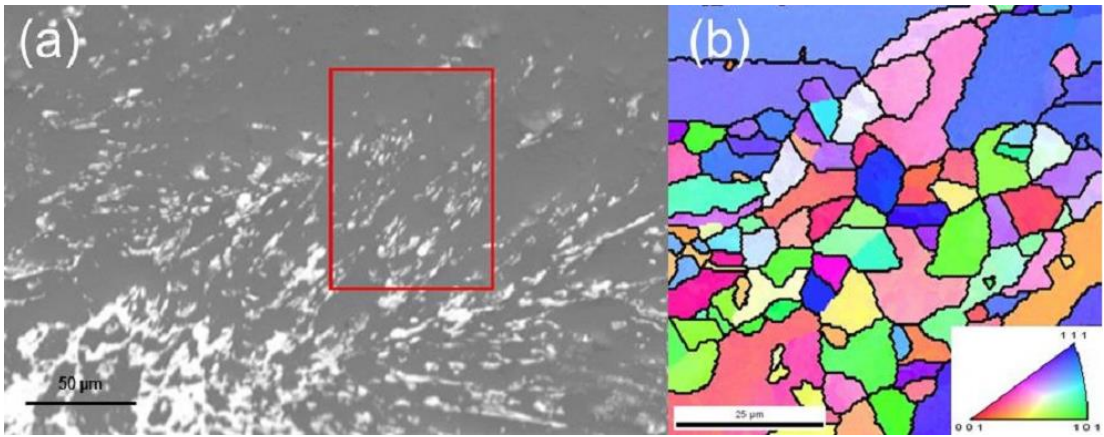

Figure 1. (a) Presence of Pd within SiC and (b) inverse pole figure (IPF) and the standard stereographic triangle (SST).

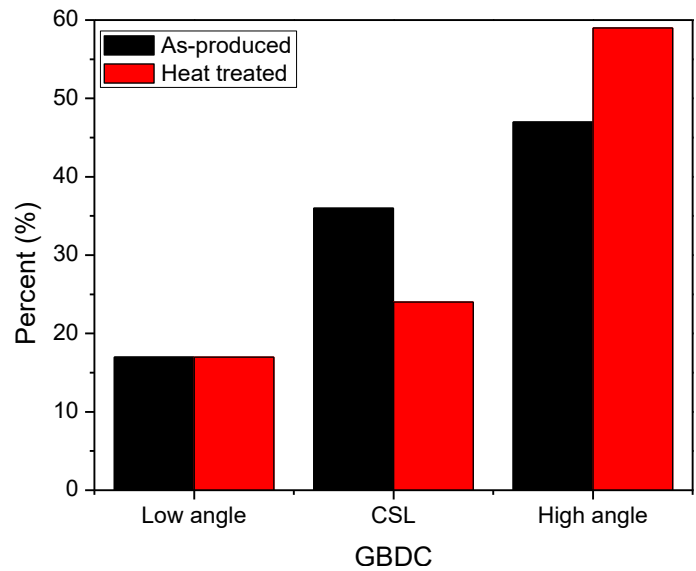

Figure 2. Grain boundary character distribution (GBCD) of SiC before and after heat treatment with Pd.

.

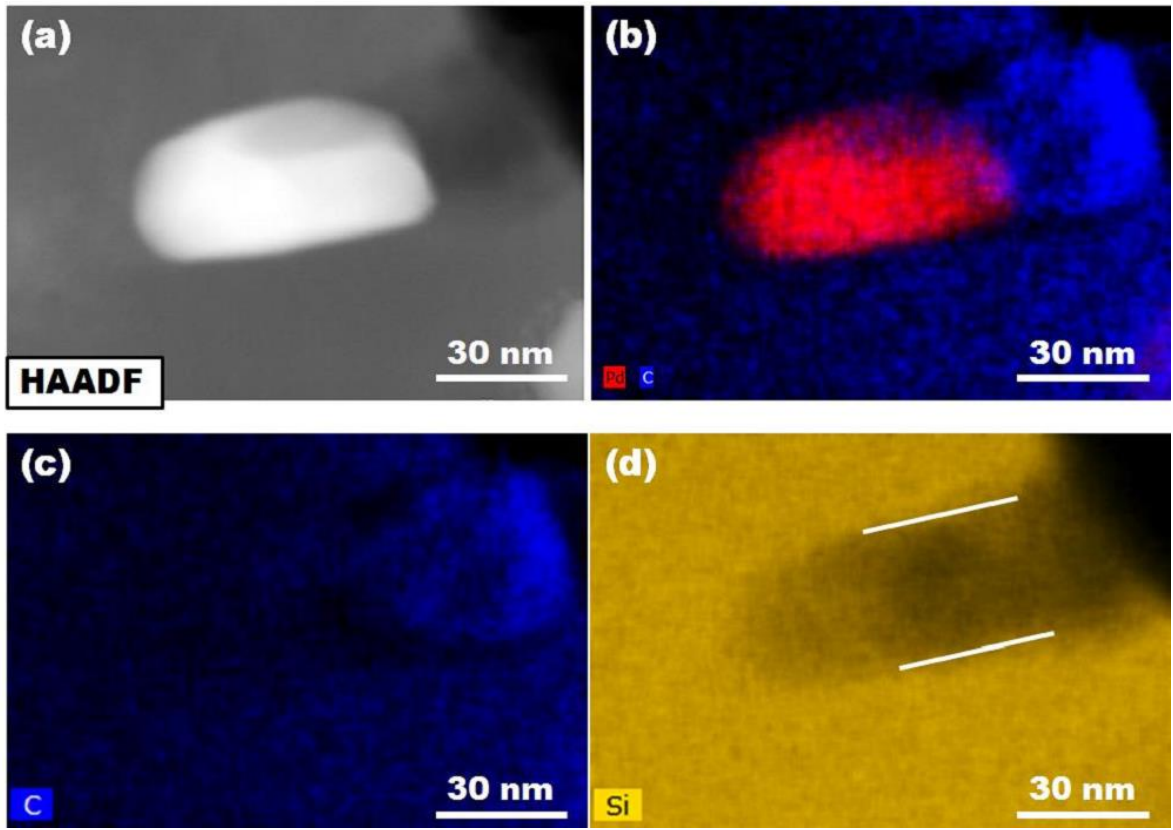

Figure 3. (a) HRTEM image of a particle of palladium in a SiC grain boundary; (b) combined Pd and C EDS map; (c) carbon EDS map y (d) silicon EDS map. In (d) is possible to observe a Si deficiency (indicated by the white bars), whereas in (c) is possible to identify the presence of excess carbon.

**Supplemental Material**

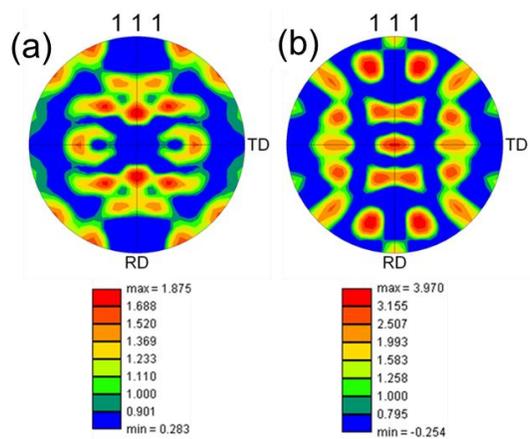

Figure S1. Pole figures of (a) as-produced SiC and (b) SiC heat treated with Pd at 1300°C.

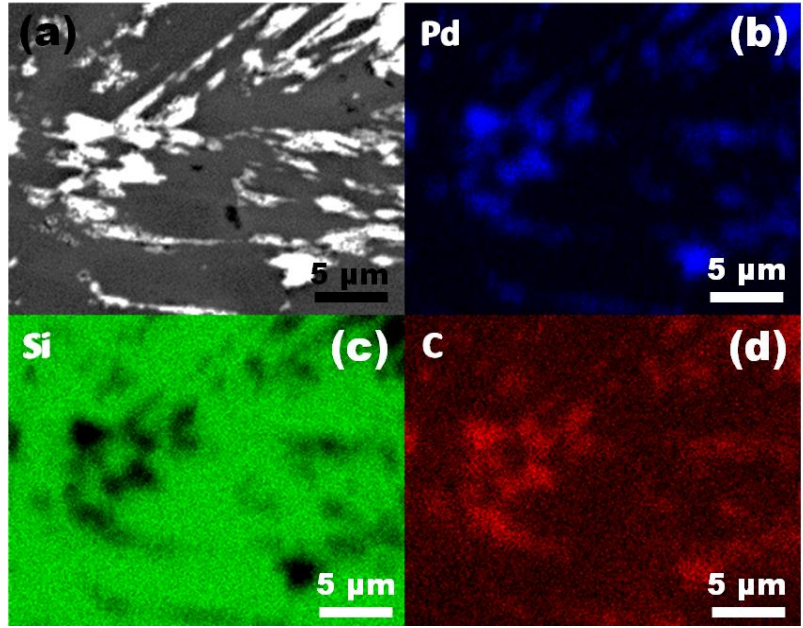

Figure S2. Element maps of an area with a high concentration of Pd in the polycrystalline monolith of SiC. (a) micrograph of Pd; (b) element map of Pd; (c) element map of Si; (d) element map of C.

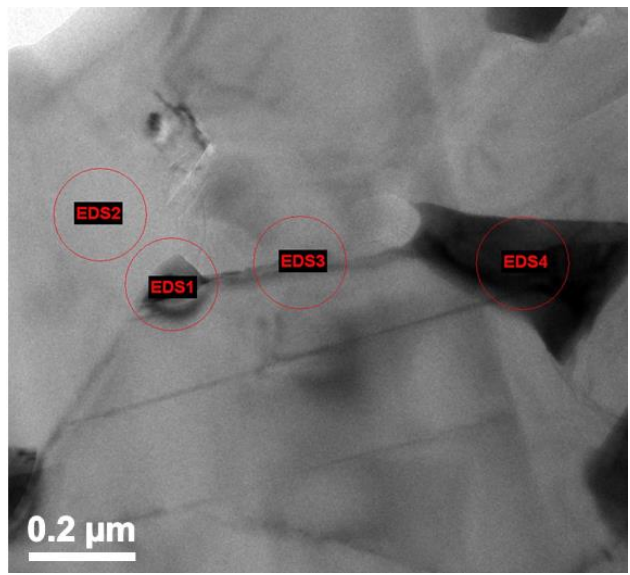

Figure S3. TEM micrograph showing the presence of two regions with Pd and the formation of a nanolayer.

Table S1. EDS element análisis of the regions described in Figure S2.

| EDS | % Atomic (C) | % Atomic (Si) | % Atomic (Pd) |
|---|---|---|---|
| 1 | 80.71 | 18.31 | 0.92 |
| 2 | 77.89 | 22.08 | 0.01 |
| 3 | 85.61 | 14.38 | 0.00 |
| 4 | 69.11 | 26.37 | 4.43 |